\documentclass[runningheads]{llncs}
\usepackage{graphicx}
\usepackage[binary-units=true]{siunitx}
\usepackage[utf8]{inputenc}
\usepackage{eurosym}
\usepackage{subfig}
\usepackage{listings}
\usepackage{xcolor}

\usepackage[hyphens]{url}
\usepackage{hyperref}
\begin{document}
	\title{Come back when you are charged!\\Self-Organized Charging for Electric Vehicles}
	\titlerunning{Come back when you are charged!}
	%
	\author{Benjamin Leiding\orcidID{0000-0002-9191-7548}}
	\authorrunning{B. Leiding}
	%
	\institute{Clausthal University of Technology\\Institute for Software and Systems Engineering\\Clausthal-Zellerfeld, Germany\\\email{benjamin.leiding@tu-clausthal.de}}
	\maketitle              

	\begin{abstract}
		
		Dwindling nonrenewable fuel reserves, progressing severe environmental pollution, and accelerating climate change require society to reevaluate existing transportation concepts. While electric vehicles (EVs) have become more popular and slowly gain widespread adoption, the corresponding battery charging infrastructures still limits EVs' use in our everyday life. This is especially true for EV owners that do not have the option to operate charging hardware, such as wall boxes, at their premises. Charging an EV without an at-home wall box is time-consuming since the owner has to drive to the charger, charge the vehicle while waiting nearby, and finally drive back home. Thus, a convenient and easy to use solution is required to overcome the issue and incentivize EVs for daily commuters. Therefore, we propose an ecosystem and a service platform for (semi-)autonomous electric vehicles that allow them to utilize their ``free"-time, e.g., at night, to access public and private charging infrastructure, charge their batteries, and get back home before the owner needs the car again. To do so, we utilize the concept of the Machine-to-Everything Economy (M2X Economy) and outline a decentralized ecosystem for smart machines that transact, interact and collaborate via blockchain-based smart contracts to enable a convenient battery charging marketplace for (semi-)autonomous EVs. 
		
		\keywords{Electric Vehicles \and Blockchain Technology \and Smart Contracts \and Internet of Things \and Circular Economy.}
	\end{abstract}

	
	\section{Introduction}
		\label{s:introduction}
		
		The catastrophic worsening of climate change, severe environmental pollution, and dwindling nonrenewable fuel reserves force society to abandon -- or at least to minimize -- fossil fuel-based transportation. Especially daily commuters burn large amounts of gas every year to drive to their workplace and back home. Electric vehicles (EVs) -- preferably powered with solar energy or other regenerative energy sources -- provide an alternative. However, even though EVs' prices are declining and closing in on gasoline-powered vehicles, they are still expensive in comparison -- despite the governmental financial promotion of EVs, e.g., in Germany~\cite{umweltbonus}.
		
		Moreover, a limited battery charging infrastructure further constraints the use of EVs in our everyday life~\cite{she2017barriers}. EV drivers who own houses have the opportunity to install the corresponding charging infrastructure, e.g., wall boxes, on their premise (and even install solar panels to power the charger), thereby circumventing the problem of limited charging infrastructure -- at least in the context of daily commuting. Charging an EV without an at-home wall box is time-consuming since the owner has to drive to the charger, charge the vehicle while waiting, and finally drive back home. This issue applies especially to people living in apartment buildings without underground parking who have to park their EVs on public streets that usually do not have an option to charge vehicles overnight. Even with a growing charging infrastructure, such parking spots will continue to lack sufficient, easy-to-use, and convenient access to charge EVs on public roads. As a result, potential buyers are disincentivized to purchase an EV not only due to the higher price but also due to the additional time-requirements and complexity of EV charging and the experimental nature of EVs. A convenient and easy to use solution is required to overcome the issue and incentivize the use of EVs for a daily commute.
		
		This work proposes a solution for (semi-)autonomous electric vehicles that allows them to utilize their ``free"-time, e.g., overnight, to access public and private charging infrastructure, charge their batteries and get back home before the owner needs the car to drive to work. To do so, we utilize the concept of the Machine-to-Everything Economy (M2X Economy) as suggested in~\cite{leidingPhD}. We present a decentralized ecosystem for smart machines that transact, interact and collaborate via blockchain-based smart contracts with other machines (Machine-to-Machine -- M2M), humans (Machine-to-Human -- M2H), and infrastructure components (Machine-to-Infrastructure -- M2I) -- thereby joining the M2X Economy -- to enable a convenient battery charging market place for (semi-)autonomous EVs. Finally, we outline how the presented solution incentivizes private wall-box owners to participate in this ecosystem, thereby fostering a private and decentralized P2P EV charging market in addition to the commercial charging infrastructure operators.		
		
		
		
		The remainder of this paper is structured as follows: Section~\ref{s:section-2} sets the idea of self-charging vehicles in the context of the M2X Economy, introduces related works as well as supplementary literature. Section~\ref{s:self-charging-evs} focuses on the operational details of our solutions. Next, Section~\ref{s:sys-engagement-processes} details selected system-engagement processes, while Section~\ref{s:discussion} discusses the findings of our paper. Finally, Section~\ref{s:coinclusion-and-future-work} concludes this work and provides an outlook on future work.
	
	
	\section{Preliminaries and Related Work}
		\label{s:section-2}
		
		Subsequent Section~\ref{ss:m2x-economy} details the concept of the M2X Economy, while Section~\ref{ss:blockchain-tech} provides background information on blockchain technology. Section~\ref{ss:related-work} outlines related work.
		
		
		\subsection{The M2X Economy}
			\label{ss:m2x-economy}
			
			The development and adoption of EVs and the development of autonomous self-driving vehicles are not necessarily correlated. However, EVs are often used to showcase the latest technological advancements in this field, e.g., Tesla. Nevertheless, no production-ready fully autonomous vehicles exist yet. The idea of fully autonomous vehicles in combination with the ubiquitousness of smart devices in general, the continuous growth and expansion of the Internet of Things (IoT)~\cite{iotGrowth2,iotGrowth1} as well as the progressing digitalization of our daily life, e.g.,~\cite{horst2013digital,su2011smart}, new technical and economical business models. While nowadays, business transactions almost exclusively focus on human-to-human transactions, the IoT enables business transactions without human intervention via autonomously acting smart machines, a concept that we refer to as the Machine-to-Machine (M2M) Economy. Besides M2M interactions, machines interact with humans (Machine-to-Human -- M2H) or infrastructure components (Machine-to-Infrastructure -- M2I). The Machine-to-Everything Economy~\cite{leidingPhD} is the result of business interactions, transactions, and collaborations among participants of the corresponding ecosystem. It represents a more general view on use cases that involve autonomous smart devices while also encompassing M2M, M2H, and M2I scenarios.
			
			In the M2X Economy, smart sensors may offer collected sensor data such as temperature or air contamination to interested buyers that rely on the aforementioned data for their own computations. In the context of autonomous and self-driving vehicles, scenarios such as automated tollbooth payments, autonomous battery charging services, as well as general Transportation-as-a-Service (TaaS) applications~\cite{leiding2018enabling} and business models are among the most discussed use cases. More complex scenarios focus on Smart Homes and Smart Cities~\cite{lynggaard2016complex} and Industry 4.0~\cite{vaidya2018industry}. Even potential successors of Industry 4.0 with fully automated and autonomous smart factories that independently handle supply and demand management and corresponding logistics -- including supply-chain management -- are part of the M2X Economy.
			
			
			Besides the technical perspective, the upcoming M2X Economy also poses sociotechnical problems and challenges. In an M2X scenario, we are not only enabling interactions, transactions, and collaborations among machines, or between machines and infrastructure components, but also among machines and humans. One of the main requirements is integrating humans and smart devices into a well-functioning sociotechnical system that puts the M2X concept in a human-centered context. When considering collaborations, interactions, and transactions of autonomous smart devices, even M2M and M2I can be seen in a sociotechnical context similar to humans interacting with each other or humans interacting with machines of the M2X ecosystem. In order to provide non-trivial services or products, smart devices are not only required to interact with their potential clients; they may also have to collaborate, interact and transact on-demand with other entities to be able to achieve a shared goal. While providing services or products, they might even migrate to various geographical locations based on supply and demand. The interleaved on-demand collaborations, interactions, and transactions among autonomous, heterogeneous, and highly dynamic entities (humans, machines, software agents, etc.) lead to a decentralized, distributed, and heterogeneous socio-technical system consisting of a large number of micro-services of different vendors and solution as well as infrastructure providers.	
			
			
			
			
		
		\subsection{Digital Smart Contracts}
			\label{ss:blockchain-tech}
			
			
			An alternative approach to the traditional oral or paper written contracts for transactions and collaborations are electronic blockchain-based smart contracts that allow to govern business transactions using a computerized transaction protocol such as a blockchain. A blockchain~\cite{nakamoto_bitcoin:2008} consists of a sequentially ordered number of blocks that records transaction events, e.g., transfer of a currency from one person to another. Transactions are cryptographically signed and represent an incremental, consistent over the network, time-stamped, and verifiable list of records~\cite{nakamoto_bitcoin:2008}. Hashes of their previous ancestor block link blocks, thereby chaining all blocks together. As a result, changing information in one block results in a hash mismatch of the succeeding block. Thus, tampering with one block requires the recalculation of all succeeding blocks. Blocks are publicly available and synchronized via a global, distributed, and decentralized P2P storage system.
			
			Furthermore, blockchain-based smart contracts allow for the automated, globally-available orchestration and choreography of heterogeneous socio-technical systems with a loosely coupled, P2P-like network structure. Finally, blockchain-based smart contract-driven platforms enable fact tracking, non-repudiation, auditability, and tamper-resistant storage of information among distributed participants without a central authority.
			
		
		\subsection{Related Works}
			\label{ss:related-work}
			
			The growing popularity of EVs results in increased research interest in EV applications and/or solutions to challenges that prevent their widespread adoption, e.g., \cite{leiding2018enabling} outlines a platform- and manufacturer-agnostic Vehicle-to-Everything (V2X) interaction and transaction layer for a vehicle-based economy. \cite{leidingPhD} expands this research to a general M2X Economy for autonomous smart devices. Both works focus on an abstract layer that enables a multitude of application scenarios within the V2X, or M2X space. However, this work extends~\cite{leidingPhD,leiding2018enabling} by instantiating an explicit application for EV charging based on their work.
			
			Other research focuses on a blockchain-supported autonomous selection of charging stations~\cite{pustivsek2016blockchain} or planning of fleet sizing and charging for autonomous EVs~\cite{zhang2019joint}.  Two recently started research projects explore the idea of charging driving vehicles via induction~\cite{induktion1,induktion2,eChargeTUBS} -- both have to be considered early-stage research. Moreover, decentralized sustainable energy markets have been considered in~\cite{mengelkamp2018blockchain,musleh2019blockchain} as well as~\cite{liu2018adaptive} in great depth. Nevertheless, they do not consider issues addressed in this work such as the utilization of free charging infrastructure components at times when the EVs are not use by their owners as well as the integration of EV charging into the a machine-driven M2X Economy.
			
			Besides academic research, various private companies prototype alternative charging infrastructure components. Volkswagen presented a prototype of a mobile charging robot for underground parking facilities with limited charging capacity~\cite{vwLadeRobot}. Tesla~\cite{teslaSnake} focuses on a snake-like charging connector that automatically connects to Tesla EVs approaching a charging station, charges the EV, disconnects once the battery is sufficiently charged, and enables the EV to empty the charging spot without the driver being presented. As a result, the unnecessary blocking of charging stations can be minimized. The German company Ubitricity~\cite{ubiCharge} offers self-developed charging kits that can be installed in standard streetlights, thereby utilizing existing power supply and converting them into EV charging stations. Except for the Ubitricity kit, all of the solutions presented above are still early prototypes and not ready for production yet. None of the presented solutions single-handedly solves the issues of limited and fixed charging infrastructures for EV owners without direct access to charging stations at their own premise or close to their home. However, they are either complementary or compatible with our proposed solution, as outlined in subsequent sections.
			
			While Ubitricity is already available in some European cities, it suffers from similar issues as standard charging stations, e.g., EVs parked on public roads without streetlights still lack sufficient, easy-to-use, and convenient access to charging points.
			
			Finally, most recent EVs, e.g., the Hyundai Kona Electric~\cite{hyundaiKona}, are already equipped with timing mechanisms to charge the vehicle in pre-defined time-slots. However, the mechanism requires the EV to be connected with a charging station.
			
	
	
	\section{Self-Charging Electric Vehicles}
		\label{s:self-charging-evs}
		
		While the idea of bringing the charging station to the EV~\cite{vwLadeRobot} -- as presented in Section~\ref{ss:related-work} -- is tempting, it lacks scalability and involves complex logistics. A sufficiently sized mobile battery to charge multiple EVs is difficult to handle. It requires a lot of space and might be dangerous to maneuver through a city due to its high energy capacity. However, small battery packs that cannot even charge a single EV or only one EV require the mobile battery pack to drive back to the charging station after every single use. Thus, instead of guiding the mobile battery pack to a nearby charging station, we propose a self-organized ecosystem of distributed, private, and/or public charging stations that routes the EV to a nearby charging system when the owner does not need the EVs services.
		
		
		\subsection{Overview}
			\label{ss:overview}
			
\begin{figure}[h]
				\centering
				\includegraphics[width=0.95\linewidth]{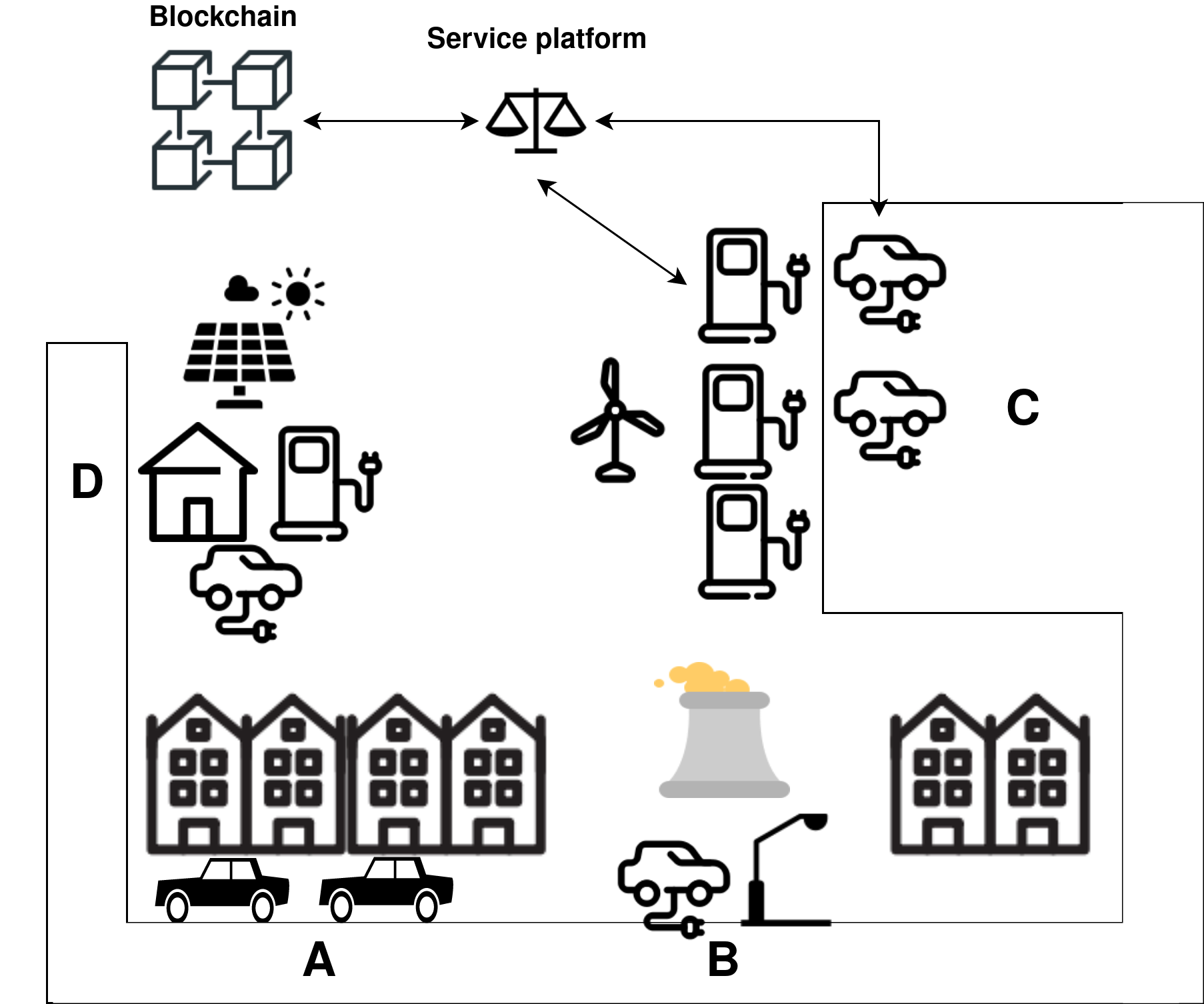}
				\caption{Self-charging EV ecosystem overview.}
				\label{fig:overview}
			\end{figure}

			Figure~\ref{fig:overview} presents the running case that we use throughout the paper to illustrate the initial problem as well as various aspects of our service platform and its ecosystem\footnote{For easier readability we refrained from adding arrows between all charging stations and the service platform.}. The running case assumes that Alice lives in an apartment building at location~\textit{A}. Due to missing underground parking or dedicated parking spots, she has to park her EV just right in front of the building along the street with no EV charging opportunities available. However, locations \textit{B}, \textit{C} and \textit{D} offer different kinds of EV charging in her neighborhood. Location \textit{B} is the closest and represents a public Ubitricity-like streetlight charging station powered by a coal-fired power plant. 
			Location~\textit{D} is also solar-powered but further away and owned by the private individual Bob who is willing to let other EV owners utilize his charging station when he does not use it himself. Finally, location~\textit{C} is even further away than location~\textit{D} but offers multiple, publicly-available, wind-powered charging stations. As a result, Alice can choose from three different neighborhood charging options in various locations and different charging properties.
			
			On-boarding private individuals such as Bob not only allows for distributed P2P electricity trading but also increases the number of available charging stations. Furthermore, Bob is incentivized to do so by selling his solar energy surplus to EV owners. Even without a surplus, he could sell power from the power grid and add an extra fee for parking, providing the charging infrastructure and compensating for his efforts.
			
			Alice commutes five days a week from her apartment to her workplace and needs to charge the EV twice a week. Due to her busy schedule, it would be great to do so overnight in such a manner that her EV is ready right before she leaves for work. Therefore, instead of relying on Alice to actively be involved in the charging process, Alices' EV utilizes its ``free" time to collaborate with the available charging infrastructure to charge itself while Alice is asleep.		
		
		
		\subsection{Coordination and Routing}
			\label{ss:routing}		
			
			Guiding the EV from Alice apartment parking spot at location~\textit{A} to any of the charging stations and back is essential to the proposed solution. We briefly outline different approaches for routing and coordination of EVs since this work's contribution focuses on such a system's system engagement processes rather than the engineering aspects.
			
			Assuming the existence and widespread adoption of fully autonomous EVs would be the easiest solution to navigate any EV to an appropriate charging station. However, this will not be the case for at least another decade. Therefore, semi-autonomous EVs that can navigate on their own in constrained scenarios, e.g., highways or slow driving in specific areas~\cite{tallinBus} might be an option. Alternatively,~\cite{induktion1} describes an approach to charge EVs while driving via induction coils. Besides charging, the coils also guide the EV. A similar approach of beacons (e.g., integrated into the street, traffic lights, streetlights, etc.) or other guiding mechanisms for low-speed, urban driving could be used in our scenario. Finally, platooning~\cite{leiding2016self} could be another option to lead groups of EVs to locations with multiple charging stations. The platoon leader could be one of the few fully autonomous EVs that might be available in the near future before widespread adoption is reached. Charging infrastructure providers may also utilize small autonomous machines that act as platoon leaders, roam the streets at night, and guide groups of EVs to their less frequented charging stations. As a result, less complex and futuristic EV hardware is required compared to fully autonomous navigation.
		
	

	\section{System-Engagement Processes}
		\label{s:sys-engagement-processes}
		
		The previous Section~\ref{s:self-charging-evs} presents the running case, describes the ecosystem, and outlines potential routing mechanisms for EVs guiding them to charging stations. However, thus far, we spared to answer how to decide which charging station to choose, which constraints apply, and how to organize the process in a self-organized, automated, and distributed manner with minimal human involvement? A core element of the use case presented above is a smart contract-based negotiation and contract enactment among entities resulting from collaborating tasks and sub-processes, e.g., an EV and a set of charging stations negotiating the terms and conditions of a battery charge. This process involves payment processing, local and global communication, and local match-making between the vehicle and the charging infrastructure. On an abstract level, battery charging negotiations are similar to many other M2X-related business enactments, e.g., road-space negotiations or parking spot fees. Therefore, we propose a specific implementation of an abstract interaction, transaction, and collaboration lifecycle that was developed for business enactments among autonomous smart devices~\cite{leidingPhD,leiding2018enabling}.
		
		In the following, Section~\ref{ss:lifecycle} details our EV charging instantiation of the abstract transaction and negotiation contract lifecycle, followed by Section~\ref{ss:auction} that details an efficient auction mechanism for M2X use cases.

		
		\subsection{Digital Contract Transaction and Collaboration Lifecylce}
			\label{ss:lifecycle}
			
\begin{figure}
				\centering
				\includegraphics[width=0.95\linewidth]{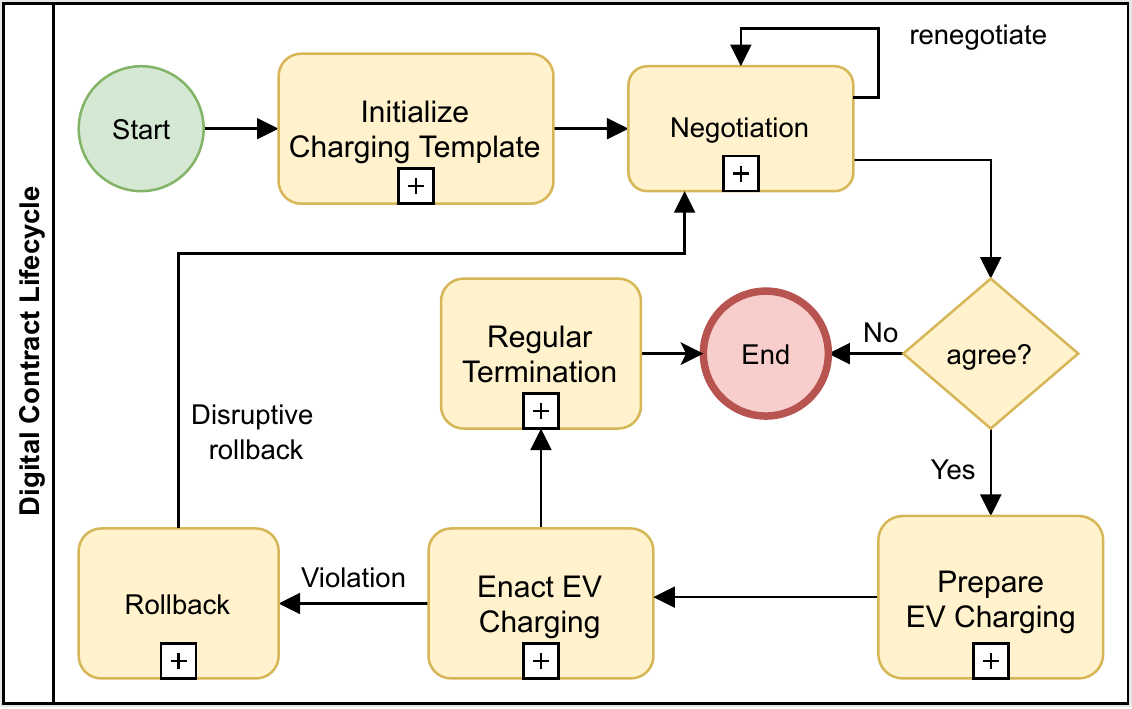}
				\caption{Digital contract negotiation and transaction lifecycle -- Based on~\cite{leidingPhD,norta2015creation,norta2015establishing,norta2016designing,norta2015conflict}}
				\label{fig:collaboration-lifecycle}
			\end{figure}

			
			In~\cite{leidingPhD,norta2015creation,norta2015establishing,norta2016designing}, the authors introduces a conceptual digital smart contract based transaction and negotiation lifecycle as illustrated in Figure~\ref{fig:collaboration-lifecycle}. The abstract nature of the proposed conceptual lifecycle allows the utilization of the same approach for various M2X-related services that rely on any kind of interactions, transactions and collaborations for M2X business enactments. The instantiated lifecycle for EV charging, as illustrated in Figure~\ref{fig:collaboration-lifecycle}, is divided into the following stages: $i.)$ initialization, $ii.)$ negotiation, $iii.)$ preparation of charging enactment $iv.)$ EV charging enactment $v.)$ rollback, and $vi.)$ termination.				
			
			During the initialization stage, information regarding the involved entities, such as identifiers and wallet addresses, are incorporated into the contract based on pre-configured templates. Additionally, the conditions of the requested contract are formally defined by specifying, e.g., the content and target of the contract. In the context of EV charging, this includes the charging station location, the source of power for charging the EV (e.g., nuclear vs. solar), the charging timeframe, charging speed, expected charging capacity, and the price per kWh. However, the requested charging contract conditions mainly depend on information such as the distance to the charger, price per kWh, and the available charging timeframe. In case the charging station and the EV agree on the negotiated conditions in stage $ii).$, they digitally sign the contract and express their approval. If no agreement is reached, a contract rollback is triggered. 
			
			In stage $iii.)$, a smart contract between the two parties is established and serves as a coordinating agent. Each participating entity receives a local contract copy containing the respective obligations of each party~\cite{norta2016designing}, e.g., charging the EV, providing sufficient energy to charge the battery, and charging speed. The participants ``obligations are observed by monitors and assigned business-network model agents (BNMA) that connect to IoT-sensors"~\cite{Qtum-WP} such as the EV's GPS-sensor. Moreover, the required process endpoints (e.g., payment processing) are provided and prepared. ``Once the e-governance infrastructure is set up, technically realizing the behavior in the local copies of the contracts requires concrete local electronic services. After picking these services follows the creation of communication endpoints so that the partners' services can communicate with each other. The final step of the preparation is a liveness check of the channel-connected services"~\cite{norta2016designing}.
			
			Afterward, in stage $iv.)$, the contract execution phase is triggered, the charging station connects to the EV and starts charging the EV's battery. The contract terminates or expires either after the EV is charged or when the contract is prematurely terminated. Failing to charge the EV as agreed upon might result in an immediate rollback of the smart contract or invokes a mediation process that is supervised by a conflict resolution escrow service that is not depicted in Figure~\ref{fig:collaboration-lifecycle}.
			
			Besides, the presented lifecycle includes incentives in case the stakeholders behave correctly as well as penalization of bad behavior, e.g., by paying a penalty. A severe violation of the contract from any of the involved parties might result in early termination or a rollback. While some conflicts may be handled in a calming manner that allows continuing the business enactment, others may cause early termination. 
			
			The lifecycle in Figure~\ref{fig:collaboration-lifecycle} presents the specific instantiation of the abstract M2X-focused lifecycle for EV charging. Collaborations for platooning and other types of M2X service enactments could be easily implemented using the lifecycle template and follow the same structure. 
		
		
		\subsection{Pricing and Negotiations}
			\label{ss:auction}
			
			Existing charging infrastructure providers already introduced a variety of pricing mechanisms for EVs. Usually, users pay a fixed price per kWh (kilowatt-hour) that varies depending on the charging speed -- faster charging is more expensive. Other pricing mechanisms combine flat rate mechanisms with additional fixed fees. An open, interoperable, automated, and machine-focused charging infrastructure as described above allows for more complex and tailored pricing mechanisms. Besides a fixed price per kWh or a flat rate, pricing structures may include current and predicted demand for given timeframes, weather forecasts -- sunny weather implies solar power availability. In contrast, windy weather increases wind power plant output, thereby reducing energy prices. The opposite is true for weather conditions with no sunshine or wind. Moreover, a charging stations' utilization also depends on its location and the time of the day, e.g., highway chargers might be highly-frequented throughout the day but almost empty overnight. Thus, it might be worth incentivizing EVs to drive a longer distance to a highway charger in exchange for a lower price at night. Finally, pricing also depends on the constraints set by the EV owner, e.g., the maximum price paid per kWh, the maximum distance the EV is allowed to travel, timeframe for charging, type of energy used to power the charging station (solar, wind, nuclear, coal, etc.).
			
			For our service platform, we utilize the benefits of fully automated machine-driven negotiations with no human-intervention. When trading goods and services, the buying and the selling party usually have contrary goals in terms of pricing. The seller's goal is to maximize profits while the buyer tries to minimize the costs. Auctions are a common approach to reach a consensus on a price between buyer and seller, but auctions among humans or humans and machines are tedious and slow. However, machine-driven auctions do not require a human-in-the-loop and conduct various auctions and many rounds in parallel. Therefore, we rely on an auction algorithm based on the concept of Vickrey auctions~\cite{moldovanu1998goethe,vickrey1961counterspeculation} that we developed for the M2X economy~\cite{leidingPhD,leiding2018enabling}. 
			
			In a Vickrey auction, participants exchange sealed (encrypted) bids, where each bidder submits a written and signed bid without having any knowledge of the bids of the other participants. After submitting all bids, the sealed bids are opened, and subsequently, the highest bidder wins. But instead of paying the price of the highest offer, the price paid is of the second-highest bid. Due to this paper's technical nature, we refer the interested reader to supplementary literature that covers additional economic and game-theoretical implications of Vickrey auctions, e.g.,~\cite{ausubel2006lovely,edelman2007internet,lucking2000vickrey,moldovanu1998goethe,vickrey1961counterspeculation}. 
			
\begin{figure}[h]
				\centering
				\includegraphics[width=\columnwidth]{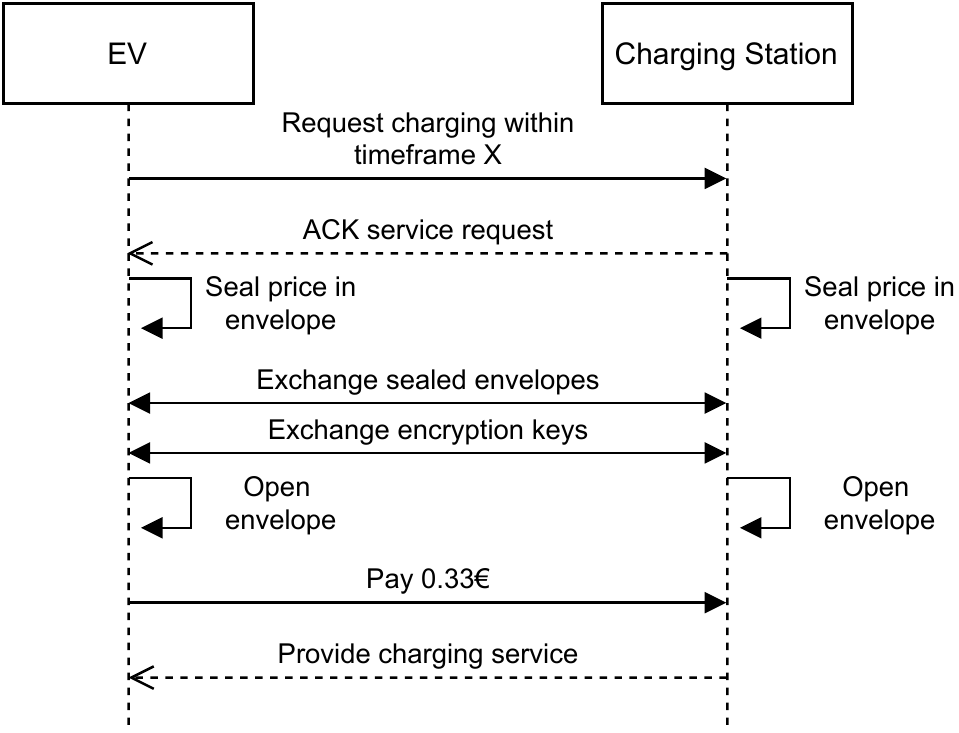}
				\caption{One-to-one Vickrey auction for EV charging -- Based on~\cite{leidingPhD,leiding2018enabling}}
				\label{fig:auction}
			\end{figure}

			Figure~\ref{fig:auction} presents the sequence diagrams of the auction algorithms that is executed during the negotiation stage of Figure~\ref{fig:collaboration-lifecycle}. Auctions may occur either locally (off-chain) between auction participants that reside close to each other or on-chain when interacting on a global scale. Only one auction round is conducted in order to minimize the runtime of the algorithm. 
			
			The auction in Figure~\ref{fig:auction} illustrates a simplified example that only focuses on the price per kWh for the following charging process and assumes that any further constraints (location, time, required charging time,fairness, etc.) have been agreed upon earlier. In case of a one-to-one auction with only one buyer and one seller, we assume that the buying EV is not willing to pay more than \EUR{0.35} per kWh (slow charging), and the charging station is not selling for less than \EUR{0.33}. Both entities prepare a sealed (encrypted) and signed bid before exchanging the bids. As soon as both parties received the other party's bid, the encryption keys are exchanged as well. EV and charging station decrypt the bids and compare the offers. Given the case that the EV offered more than \EUR{0.33}, the auction is successful, and due to the second-price rule of Vickrey auctions, the buyer pays \EUR{0.33} to the seller. In case the EV offers less than the charging station's minimum price, the auction ends without an agreement. A scenario with multiple buyers, e.g., platooning vehicles, works analogously. The auction algorithm even supports scenarios with multiple sellers, i.e., competing charging stations and scenarios with multiple buyers and sellers. In the end, the highest bidding EV is paying the second-highest price to the charging station with the highest minimum price, and so on -- as long as the paid price is higher than the matched seller's minimum price.		
		
		

	\section{Discussion}
		\label{s:discussion}
		
		The presented ecosystem and the charging service platform predominantly focus on solving the issue of limited and fixed charging infrastructures for EV owners without direct access to charging stations at their own premises or in close proximity to their home. 
		
		The service platform, its smart contracts, the auction mechanism, and the digital contract lifecycle management merely require additional software and communication-interface upgrades. However, the corresponding hardware upgrades for the EVs and the routing-related smart city components, i.e., beacons for precise navigation, may require more time, upfront investments, and further research. Waiting for fully autonomous EV navigation is also not a short-term option and rather an additional feature for future EV generations. 
		
		Further limitations arise from the required interoperability among EVs and charging stations. A one-stop platform is desirable but not a manufacturer-focused platform with deliberately forced or functional lock-ins. As suggested in~\cite{leidingPhD}, interoperability allows for the exploitation of economies of scale and increased efficiency. Moreover, a smart contract-driven service platform and its corresponding ecosystem, as described in previous sections, not only ``enable an interoperable platform for autonomous smart devices (e.g., vehicles), but also further reduces dependency on intermediaries. Furthermore, a blockchain-based solution enables the decentralized settlement of value added in the form of crypto tokens; these can be created entirely without central instances or intermediaries and exchanged directly P2P~\cite{cap2018blogchain}" while at the same time increasing transaction speed~\cite{leidingPhD}.
		
		Finally, legal concerns may apply in the context of machine-driven business enactments with no or minimal human supervision regarding liability and the legally binding nature of contracts among machines.	
	
		
	\section{Conclusion and Future Work}
		\label{s:coinclusion-and-future-work}
		
		This paper introduces a transaction and service platform for distributed EV charging and its corresponding ecosystem to overcome the issue of limited fixed charging infrastructures for EV owners without direct access to charging stations at their own premises or in close proximity to their home. We introduce a service platform for (semi-)autonomous EVs that allows them to utilize their ``free"-time, e.g., at night, to access public and private charging infrastructure, charge their batteries, and get back home before the owner needs the EV again. To do so, we utilize the concept of the Machine-to-Everything Economy and outline a decentralized ecosystem for autonomously acting smart devices that transact, interact and collaborate via blockchain-based smart contracts to enable a convenient battery charging market place.
		
		First, we present a decentralized ecosystem and our service platform for (semi-)autonomous EVs that transact, interact, and collaborate via blockchain-based smart contracts with other machines (Machine-to-Machine -- M2M), humans (Machine-to-Human -- M2H), and infrastructure components (Machine-to-Infrastructure -- M2I) -- thereby joining the M2X Economy -- to enable a convenient battery charging marketplace. The ecosystem incorporates charging stations owned by individuals and businesses as well as public charging stations. The blockchain enables smart contract-based matchmaking and further monitors and tracks all contract-related activities.	
		
		Digital smart contracts allow for a high degree of self-organization and automation among EVs. The smart contract-driven service platform offers matchmaking services among charging stations and EVs and enables smart contract-based transactions and collaborations using a digital contract lifecycle management. Moreover, it uses pre-defined contract templates in combination with the contract lifecycle management to cover the whole process of routing vehicles to the charging station as well as all charging-related business enactments in an automated manner -- including conflict-resolution mechanisms. Moreover, smart contracts ensure fact tracking, non-repudiation, auditability, and tamper-resistant storage of information among all entities and orchestrate related heterogeneous software services of involved stakeholders. 
		
		Pricing and negotiations among stakeholders rely on a Vickrey auction algorithm for the M2X Economy that allows reaching an efficient consensus on a price between EV and charging station. The algorithm is suitable for one-to-one auctions, one-to-many auctions as well as many-to-many auctions. Finally, the presented solution incentivizes private wall box owners to participate in this ecosystem and sell free charging capacities to EV owners without a wall box on their own premise, thereby fostering a private and decentralized P2P EV charging market in addition to the commercial charging infrastructure operators.	
		
		Future work focuses on a prototype implementation of the smart contract-driven service platform and defining contract templates for charging and routing-related business enactments. Moreover, we will test the proposed solution in a testbed or lab-environment with small UAVs. Finally, we will explore the applicability of the proposed solution beyond EVs to UAVs, e.g., drones or other (semi-)autonomous smart devices that fall within the context of the M2X Economy.
	
	
	
	
	%
	%
	%
	 \bibliographystyle{splncs04}
	 \bibliography{bibliography}

\end{document}